# A Polarized Positron Source for CEBAF

J. Dumas[a,b], J. Grames[b], E. Voutier[a]

[a]*Laboratoire de Physique Subatomique et de Cosmologie*
*IN2P3/CNRS – Université Joseph Fourier - INP*
*53, rue des Martyrs, 38026 Grenoble Cedex, France*

[b]*Thomas Jefferson National Accelerator Facility*
*12000 Jefferson Avenue, Newport News, VA 23606, USA*

**Abstract.** A compact polarized positron source for Jefferson Lab is discussed. This scheme relies upon a high polarization (85%), high current (1 mA), low-energy (<100 MeV) electron beam to generate polarized positrons in a conversion target via polarized bremsstrahlung and pair creation. GEANT4 is used to simulate source distributions suitable for a CEBAF-like injector with positron polarization ~60% and nano-Ampere intensity. An experiment to test this scheme is outlined.

## INTRODUCTION

The nuclear physics program at the Continuous Electron Beam Accelerator Facility (CEBAF) at Jefferson Laboratory (JLab) depends upon an electron beam with high polarization (85%). Analogously, a polarized positron beam would augment the program, enhancing the precise understanding of nuclear structure [1], as investigated via the polarized beam charge asymmetry measured in elastic [2] or deeply virtual Compton (DVCS) [3,4] scattering.

Two main techniques for positron production exist: $\beta^+$-decay by radioactive sources and pair creation from a photon in an atomic field. For accelerator purposes, pair creation is preferred in order to achieve time-structured positron beams. A conventional un-polarized positron source involves a high energy (MeV-GeV) electron beam interacting with a target of high atomic number (Z), resulting in a bremsstrahlung shower with photon energies sufficient for pair creation. Notwithstanding the prominent technical issues to capture and accelerate the positrons to a beam [5], a conventional *polarized* positron source has been explored [6,7] in the context of a future linear collider. In this scheme, the polarization of a longitudinal electron beam is transferred to the resulting positrons following the calculations by Olsen & Maximon [8]. Most recently, advances in high-polarization (85%), high-current (1 mA) electron sources [9] offer even greater potential for a compact, low energy (<100 MeV) driver for a polarized positron source. This paper builds upon work [10] aiming to experimentally characterize such a polarized electron driver source of polarized positrons at CEBAF.

# POLARIZATION TRANSFER: ELECTRON TO POSITRON

The polarized cross-sections of bremsstrahlung and pair production, including Coulomb corrections and exact atomic screening, have been calculated by Olsen & Maximon [8] in the high energy limit. In these calculations, the screening corrections are derived for Thomas-Fermi atoms and depend on the parameter $\Delta$ defined as [8]:

$$\Delta = \frac{Z^{1/3}}{121} \frac{12\, E_1 E_2}{k} \frac{1}{1 + k \cdot \sin\theta} \begin{cases} \Delta < 0.5 & \text{No atomic screening} \\ 0.5 \leq \Delta < 120 & \text{Intermediate screening} \\ 120 \leq \Delta & \text{Complete screening} \end{cases}$$

Where k is the photon energy, $E_1$, $E_2$ are the energies of either the incident/scattered electron (bremsstrahlung) or electron/positron (pair creation), in units of electron mass and $\theta$ is the scattering angle of the particle created.

In the case of bremsstrahlung, the circular polarization of photons created by 60 MeV longitudinally polarized electrons scattering from tungsten is rather well-defined (Fig. 1, left). In small angle approximation assumed by these calculations, the polarization of the photons created by incident relativistic electrons depends essentially on the fractional transfer of energy ($E_\gamma/E_{e^-}$). However, the characteristic "S" shape is notably unphysical (>1) when almost all of the electron kinetic energy is transferred to a photon, that is, the relativistic approximation may be questioned at this extreme where the scattered electron is notably non-relativistic. This is better demonstrated in the case of a 3 MeV electron beam (Fig. 1, right).

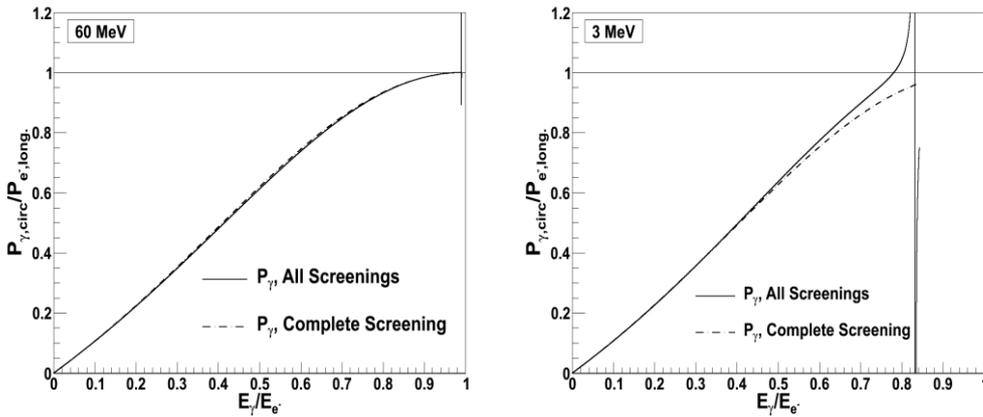

**FIGURE 1.** Circular polarization transfer of photons created by 60 MeV (left) and 3 MeV (right) longitudinally polarized electrons on a tungsten atom and different screening.

Pair production exhibits a similar pattern (Fig. 2, left) where the abscissa is now the fractional energy transferred from photon to positron ($E_{e^+}/E_\gamma$). Unphysical regions are again prominent for the low energy positron or electron. These regions are particularly relevant in a conversion target because the $1/E_\gamma$ bremsstrahlung cross-section means that most of the photons are produced at small energy and create non-relativistic electron-positron pairs. This is further demonstrated by considering low energy photons, for example, 3 MeV (Fig. 2, right).

By artificially enforcing *only* complete screening (dashed line) the polarization transfer appears continuous and manageable. This ad-hoc implementation, valid at high lepton energies, is used here as an intermediate solution for more reliable simulations. Exact calculations relaxing the high energy and small angle approximations are under investigation [11] and will soon be implemented into GEANT4.

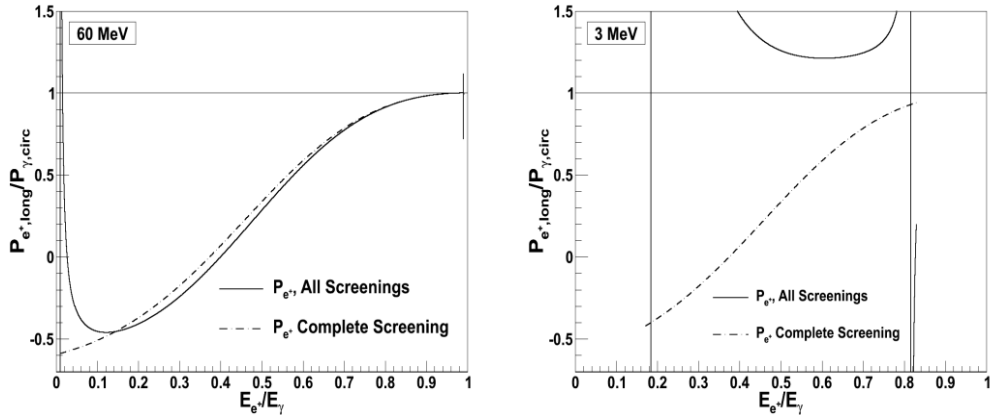

**FIGURE 2.** Longitudinal polarization transfer of positrons created by 60 MeV (left) and 3 MeV (right) circularly polarized photons interacting with a tungsten atom and different screening.

## GEANT4 SIMULATIONS FOR POLARIZED E+ SOURCE

The case for a CEBAF-like injector electron beam (E = 60 MeV, P = 85%) and a 100 μm tungsten conversion target has been simulated using GEANT4 [12]. The simulation package includes polarization observables for electromagnetic processes described in [8,13]. The positron yield (Y), average polarization (P) and figure of merit (FoM) as a function of the positron kinetic energy is shown in Fig. 3. The FoM, defined as $Y \cdot P^2$, of practical interest in experiments, is inversely proportional to the achievable statistical accuracy in a given amount of time. It is found here to occur at half of the electron beam energy.

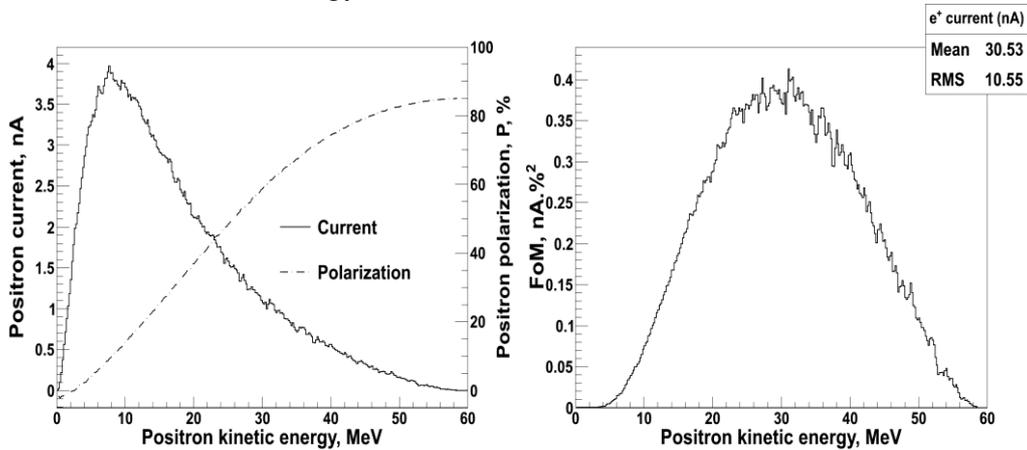

**FIGURE 3.** Positron spectra for a 1 mA and 85 % longitudinally polarized electron beam radiating in a 100 μm tungsten target: yield (left), average polarization (left), and FoM (right).

Similar simulations were performed with electron beam energies from 3 to 60 MeV. In each case the positron distribution exiting the foil was restricted to ±10° angular acceptance and binned with ±0.25 MeV energy acceptance, approximating the injector admittance. The yield and polarization at maximum FoM is plotted in Fig. 4 as a function of the electron beam energy. The positron polarization is found to be greater than ~60% for electron energies greater than ~5 MeV. In addition, the positron yield is greater than 1 nA for electron energies greater than ~20 MeV. The optimum beam energy is very sensitive to the positron acceptance cuts applied; the dependence on a positron collection system and consideration of a precise CEBAF accelerator admittance is required to make more refined statements. Additionally, different target thicknesses will be investigated to address depolarization effects and power deposition issues of the conversion target.

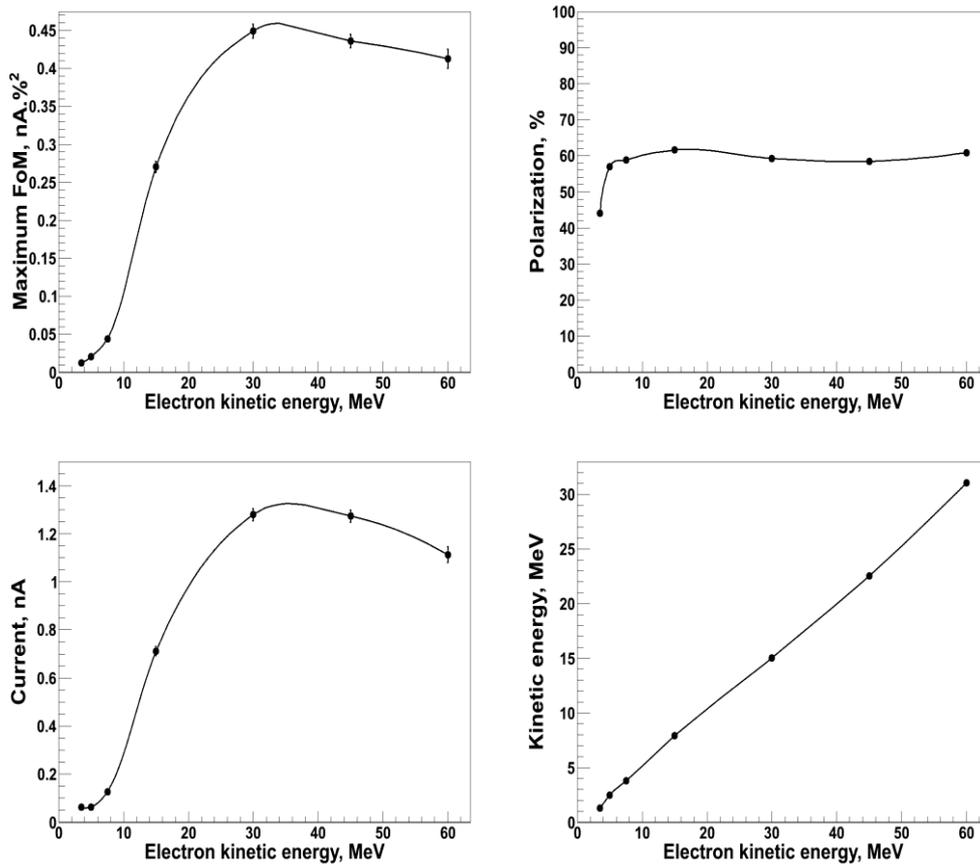

**FIGURE 4.** Positron distribution at maximum FoM as a function of the electron beam energy for a 1mA 85% longitudinally polarized electron beam and 100μm thick tungsten conversion target.

# EXPERIMENT DEMONSTRATION

The concept described above may be explored experimentally at JLab by integrating the CEBAF polarized electron source and injector as a driver for a positron source. Advances in superlattice photocathodes (GaAs/GaAsP) simultaneously achieve both high electron polarization (85%) and high quantum efficiency (~1%) when illuminated with near-infrared light (780 nm) and have produced high current (1 mA) for sustained periods with good photocathode lifetime (>200 C) in a DC high voltage electron gun [10]. CEBAF injector radio-frequency cavities then bunch and accelerate the electron beam to sufficiently high energy (presently 60 MeV, planned 120 MeV) where the electron beam emittance, energy and polarization may be measured. To test this scheme a new beam line with conversion target, positron polarimeter and existing electron beam dump is planned.

The conversion target should have a high atomic number and high melting point in order to create an electro-magnetic shower as large as possible. The beam power deposition in a production target resulting from high electron beam intensities can reach several kilo-Watts and require technological developments [5]. To avoid both technical and costly obstacles the demonstration experiment will use an electron beam with a low duty-factor (<1%) macro-pulse structure. However, the electron bunch charge will be commensurate with a 100% duty factor continuous-wave beam of 1 mA at 1.5 GHz, to assess full power production event rates. In addition, different target thickness and incident beam angle [14] will be considered to study multiple scattering, positron collection and dumping the primary electron beam.

The polarimeter diagnostics follows from the technique reported by the E166 collaboration [15], used in a demonstration experiment for the ILC polarized positron source [16]. This includes a spectrometer and focal plane Compton transmission polarimeter to characterize the positron yield, energy and polarization. A pair of identical dipoles and a collimator select the positron energy to be analyzed which, as a result of successive opposite deflections, conserves the longitudinal polarization en route to the polarimeter. There, the polarized positrons are converted in a radiator target to polarized photons which interact with a polarized target. The Compton transmission asymmetry between two opposite target polarizations is measured with CsI calorimeter. Systematic effects of the polarimeter may be studied by rapid reversal (up to 1 kHz) of the electron beam polarization (and positrons). Furthermore, polarity of the diagnostics may be reversed to instead measure the pair-produced electron polarization.

## SUMMARY & OUTLOOK

The production of spin-polarized positrons via polarized bremsstrahlung and pair creation is considered in the context of a polarized positron source for JLab. Based on the existing CEBAF polarized electron source, a polarized positron beam would support a new aspect of the existing experimental program. For example, a Deep Virtual Compton Scattering (DVCS) experiment would require a polarized positron beam intensity of ~10 nA. Calculations and GEANT4 simulations indicates nano-Ampere current with polarization of 60% should be possible. An experiment is planned to characterize the positron distribution using this scheme. This will help to benchmark modeling tools and collect valuable technical information for the design, optimization and construction of an actual source.

## ACKNOWLEDGMENTS

This work was supported in part by the U.S. Department of Energy (DOE) contract DOE-AC05-06OR23177 under which the Jefferson Science Associates, LLC, operates the Thomas Jefferson National Accelerator Facility, the French National Center of Scientific Research, and the GDR n°3034 Physique du Nucléon.